\documentclass[%
 aip,jcp,
 amsmath,amssymb,
reprint,%
]{revtex4-1}

\usepackage[utf8]{inputenc}
\usepackage{graphicx}
\usepackage{dcolumn}
\usepackage{bm}
\usepackage{mathtools}

\usepackage[T1]{fontenc}
\usepackage{mathptmx}

\newcommand{\matr}[1]{\mathbf{#1}}

\newcommand{\op}[1]{\ensuremath{\mathsf{#1}}}
\newcommand{\vect}[1]{\ensuremath{\bm{#1}}}
\renewcommand{\vec}[1]{\ensuremath{\bm{#1}}}

\newcommand{\mean}[1]{\ensuremath{\left<#1\right>}}
\newcommand{\adj}[1]{\ensuremath{#1^{\dagger}}}

\newcommand{\ts}{\ensuremath{\textrm{S}}}
\newcommand{\te}{\ensuremath{\textrm{E}}}
\newcommand{\tee}{\ensuremath{\textrm{EE}}}
\newcommand{\tse}{\ensuremath{\textrm{SE}}}
\newcommand{\tes}{\ensuremath{\textrm{ES}}}
\newcommand{\tss}{\ensuremath{\textrm{SS}}}

\newcommand{\df}{\ensuremath{\mathrm{d}}}
\newcommand{\ddt}{\ensuremath{\frac{\mathrm{d}}{\mathrm{d} t}}}

\begin{document}

\title{Context in Synthetic Biology: Memory Effects of Environments with Mono-molecular Reactions}

\author{Johannes Falk}
\email{falk@fkp.tu-darmstadt.de}
\affiliation{ Institut für Festkörperphysik, Technische Universität Darmstadt, Hochschulstr. 6, 64289 Darmstadt, Germany }%

\author{Leo Bronstein}
\affiliation{Bioinspired Communication Systems, Technische Universität Darmstadt, Rundeturmstr. 12, 64283 Darmstadt, Germany}%
\affiliation{Institute of Biostatistics and Clinical Research, University of Münster, Schmeddingstr. 56, 48149 Münster, Germany}%

\author{Maleen Hanst}%
\affiliation{Bioinspired Communication Systems, Technische Universität Darmstadt, Rundeturmstr. 12, 64283 Darmstadt, Germany}%

\author{Barbara Drossel}
\affiliation{ Institut für Festkörperphysik, Technische Universität Darmstadt, Hochschulstr. 6, 64289 Darmstadt, Germany }%

\author{Heinz Koeppl}
 \email{heinz.koeppl@bcs.tu-darmstadt.de}
\affiliation{Bioinspired Communication Systems, Technische Universität Darmstadt, Rundeturmstr. 12, 64283 Darmstadt, Germany}%

\date{\today}

\begin{abstract}
Synthetic biology aims at designing modular genetic circuits that can be assembled according to the desired function. When embedded in a cell, a circuit module becomes a small subnetwork within a larger environmental network, and its dynamics is therefore affected by potentially unknown interactions with the environment. It is well-known that the presence of the environment not only causes extrinsic noise but also memory effects, which means that the dynamics of the subnetwork is affected by its past states via a memory function that is characteristic of the environment. We study several generic scenarios for the coupling between a small module and a larger environment, with the environment consisting of a chain of mono-molecular reactions. By mapping the dynamics of this coupled system onto random walks, we are able to give exact analytical expressions for the arising memory functions. 
Hence, our results give insights into the possible types of memory functions and thereby help to better predict subnetwork dynamics.
\end{abstract}

\maketitle

\section{\label{sec:intro}Introduction}

One of the key goals of synthetic biology is the development of modular genetic devices that can be assembled in a flexible manner according to the desired task\cite{shetty_engineering_2008}. Each device is designed to provide exactly one function in a robust and reliable fashion, analogous to, e.g., resistors and capacitors in electrical circuits. Conventionally, development and optimization of each module are done in isolation, and the output parameters are analyzed under controlled conditions. Based on this characterization, complex genetic circuits of several interconnected modules are designed, optimized and implemented. However, upon interconnection and implementation into a cell, each circuit finds itself embedded in a fluctuating environment that possibly affects the intended functioning\cite{andrianantoandro_synthetic_2006}. In the field of synthetic biology these environmental influences are often subsumed as context effects and can, according to their origin, be subdivided into different categories\cite{purnick_second_2009,cardinale_contextualizing_2012}: \emph{Compositional} context describes the perturbation of a module due to the functional composition of the device, \emph{host} context captures the influences provided by the cellular environment, and \emph{external} context captures all disturbances that originate outside the cell. In order to design reliable modules, it is essential to analyze and subsequently robustify network designs specifically against these different and often unavoidable contextual effects. Consequently, several authors have studied the impact of context effects and proposed methods to account for them already during the design process of genetic devices:

Saez-Rodriguez et al.~and later Del Vecchio et al.~considered compositional context effects by investigating how the addition of a downstream module can change the output of the upstream module. This effect was coined \emph{retroactivity}\cite{saez-rodriguez_dissecting_2005,del_vecchio_modular_2008}. The authors proposed a general approach to quantifying the effects of retroactivity. Additionally, they designed an insulating device that attenuates retroactivity and later successfully demonstrated its functionality. A basic assumption for Del Vecchio's analysis is that there is a separation of time scales between the dynamics of the two described modules. While this assumption is often valid for  compositional context effects, it cannot be applied to host context effects, since the cellular environment  often has similar time scales to the designed subnetwork. In spite of this, the interactions with the host environment are often ignored by assuming that the copy numbers of environmental species are very large or the environment is efficiently buffered. An important step forward was made by Liebermeister et al.~who developed a general framework that can be used to include compositional as well as host context effects into the rate equations, especially accounting for the dynamical response of the environment to the output of the subnetwork\cite{liebermeister_biochemical_2005}. To this purpose, they divided the full network into a network of interest, the so-called subnetwork, and the surrounding environment. After linearizing the equations for the environmental variables around a steady state, they reduce its dimensionality by projecting the dynamics on the subspace of the dominant dynamic modes.
Applying their method to a metabolic network embedded in an environment with parameter uncertainty, Liebermeister et al.~demonstrated how accounting for the environment improves the predictions of the model, even if the parameters are not known exactly. While model reduction via balanced truncation was shown to outperform conventional approaches based on fixed environmental concentrations, Liebermeister et al.~also mention several limitations of their method, namely the requirement for a steady state in the environment and the non-preservation of conservation relations between subnet and environment.
The results of Liebermeister et al.~were recently extended by Rubin et al.~who used the Mori-Zwanzig projection operator formalism to obtain a description for a generic subnetwork that includes the contextual effects via so-called memory terms \cite{rubin_memory_2014}. By using memory terms, it is possible to include complex mutual interaction between subnetwork and environment into the description of the subnetwork, for instance when the subnetwork  influences the environment and these environmental changes are coupled back to the subnetwork later in time. Using the projection technique, the authors demonstrate that the inclusion of memory terms leads to higher accuracy in the predicted subnetwork dynamics.

A different approach to including environmental effects into the description of the subnet was pursued by Zechner et al.\cite{zechner2014scalable, zechner2014uncoupled} and Bronstein and Koeppl\cite{PhysRevE.97.062147}. In the description of the stochastic process for the evolution of the joint network, the environmental states were integrated out, resulting in a description of the subnet that uses the best estimate of the environmental state, based on the past observations of the subnet.

In this paper, we will extend these various investigations of environmental effects on the subnet in several respects. In contrast to Rubin et al., who based their derivation on the chemical Fokker-Planck equation (CFPE) in the limit of vanishing noise, we will apply the Nakajima-Mori-Zwanzig projection operator framework directly to the chemical master equation (CME) to marginalize out the environmental states, since the CME is the more fundamental equation for stochastic modeling of a chemical reaction network.
We will show that for a reaction network with linear environment and linear subnet-environment interactions, the marginal description of the CME still possesses closed first-order moment equations and that these coincide with the equations obtained by applying the projection framework to an ODE model of the same form as the reaction rate equation (RRE). 
While gene-regulatory networks are usually nonlinear, they can in certain situations be simplified to linear networks either by expansion around a steady-state \cite{liebermeister_biochemical_2005} or by mapping them on approximately equivalent systems \cite{cao_linear_2018}. 

Based on this linear description, a main goal of our paper is the systematic study of the influence of generic environments on subnetworks. In distinction to previous studies of marginalization\cite{rubin_memory_2014}, we focus on the situation that we do not have specific knowledge of the environment into which a module might be embedded.
By considering several chain-like prototypical environments (e.g. cascaded genetic module environments), we derive analytical expressions for the memory kernels of these common cellular environmental structures. Those environmental building blocks can be used as general-purpose environment models when investigating the sensitivity of a module to contextual influences.
To demonstrate the effects of such contextual influences, we take, similarly to Del Vecchio et al., the example of an oscillating genetic module that interacts with an environment of mono-molecular reactions. Additionally, we demonstrate a more complex context effect using the example of a negative-feedback module that drives interconnected downstream modules.
This shows that even environments that seems to be too simple to be relevant considerably change the dynamics of a module. Furthermore, 
 our analytical results can be extended to include a broad class of environments into the analysis of synthetic genetic networks.

The outline of the remainder of the paper is as follows:
In Sec.\,\ref{sec:methods} we derive stochastic and deterministic descriptions for the subnet that contain the effects of the environment via memory functions, and relate the corresponding marginal description.
Sec.\,\ref{sec:lin_environments}  presents derivations of exact analytical expressions for the memory terms of prototypical environments, which will be based on the deterministic marginal description.
Finally in Sec.\,\ref{sec:examples}, we use the memory functions of several simple model environments and demonstrate their effect on two specific subnets.

\section{\label{sec:methods}Model and methods}

\subsection{Reaction network: System and environment} \label{sec:setting}
In the following, we call the system of interest the \textit{subnetwork} or \textit{target network}, indicating that it is part of a larger network that comprises the system of interest and its environment. We model the full system as a reaction network consisting of $N+M$ species, where $N$ and $M$ denote the number of subnet and environmental species, respectively, and $R$ reactions, 
\[
\sum_{n=1}^{N}{s_{nj} \textrm{X}^{\ts}_n} + \sum_{n=1}^{M}{s'_{nj} \textrm{X}^{\te}_n} \longrightarrow \sum_{n=1}^{N}{r_{nj} \textrm{X}^{\ts}_n} + \sum_{n=1}^{M}{r'_{nj} \textrm{X}^{\te}_n}
\]
for $j = 1, \dots, R$. The superscripts S and E denote the subnet and environmental species, respectively.  The coefficients $s_{1j}, \dots, s_{Nj}$ and $s'_{1j}, \dots, s'_{Mj}$ are  the stoichiometric substrate coefficients and $r_{1j}, \dots, r_{Nj}$ and $r'_{1j}, \dots, r'_{Mj}$ the stoichiometric product coefficients of reaction $j$.

\subsection{Stochastic and deterministic dynamics} \label{sec:dynamics}

For a well mixed system 
the time evolution of the network is modeled by a continuous-time Markov chain $\vect{X}(t) = (\vect{X}^{\ts}(t), \vect{X}^{\te}(t))$ in which the state $\vect{x} = (\vect{x}^{\ts}, \vect{x}^{\te}) \in \mathbb{N}_0^{N} \times \mathbb{N}_0^{M}$ of the system specifies the (integer-valued) copy number of each species.
The probability $p(t, \vect{x}) := P(\vect{X}(t) = \vect{x})$ evolves according to the CME
\begin{equation} \label{eq:cme}
\frac{\df}{\df t}  p(t, \vect{x}) = \op{T} p(t, \vect{x})
\end{equation}
with forward evolution operator 
\begin{equation} \label{eq:cme_forward_operator}
[\op{T} p](\vect{x}) = \sum_{j=1}^R\left\{ h_j(\vect{x}-\vect{\nu}_j)p(\vect{x}-\vect{\nu}_j) -  h_j(\vect{x})p(\vect{x})\right\}.
\end{equation}
Here, $h_j(\vect{x})$ is the rate function of reaction $j$ when the system is in state $\vect{x}$, also called hazard function, and $\vect{\nu}_j = (\vect{\nu}^{\ts}_j, \vect{\nu}^{\te}_j)$ is the stoichiometric change vector of reaction $j$ with $\vect{\nu}^{\ts}_j = (r_{1j} - s_{1j}, \dots, r_{N j} - s_{N j})$ and $\vect{\nu}^{\te}_j = (r'_{1j} - s'_{1j}, \dots, r'_{M j} - s'_{M j})$. 

We say that a stochastic reaction network has \emph{mass-action kinetics} if its hazards are given by 
\[
h_j(\vect{x}) = \tilde{c}_j \prod_{i=1}^{N} \binom{\vect{x}_i^{\ts}}{s_{ij}} \prod_{l=1}^{M} \binom{\vect{x}_l^{\te}}{s'_{lj}}
\]
for $j=1, \dots, R$, where $\tilde{c}_j$ is a constant rate parameter.

A reaction is said to be linear if its hazard function is linear. For a system with mass-action kinetics, a reaction is linear if at most one stoichiometric substrate coefficient is nonzero, i.e., a linear reaction takes one of the two forms
\[
\emptyset \longrightarrow [\cdots], \quad \quad \textrm{X}_i \longrightarrow [\cdots],
\]
where the right-hand side of the reactions can be arbitrary. A reaction is said to be mono-molecular if at most one stoichiometric substrate as well as one stoichiometric product coefficient is nonzero, i.e., it takes one of the forms
\[
\emptyset \longrightarrow \textrm{X}_i, \quad \textrm{X}_i \longrightarrow \emptyset, \quad \textrm{X}_i \longrightarrow \textrm{X}_j,
\]
where the first reaction type is called a birth, the second a degradation, and the last a conversion reaction. Every mono-molecular reaction is linear.

Finally, a reaction network is termed linear (resp. mono-molecular) if all reactions are linear (resp. mono-molecular). In particular, we define the environment as linear (resp. mono-molecular) if all environment-environment interactions are linear (resp. mono-molecular).

Further, recall that for a function $\psi(\vect{x})$ the evolution equation for the moment $\mean{\psi}_t = \sum_{\vect{x}}p(t, \vect{x})\psi(\vect{x})$ is given by
\begin{equation}
\begin{aligned}
\ddt \mean{\psi}_t &= \sum_{\vect{x}} \psi(\vect{x}) \op{T} p(t, \vect{x}) = \sum_{\vect{x}} [\adj{\op{T}} \psi](\vect{x}) p(t,\vect{x}) \\
&= \mean{\adj{\op{T}} \psi}_t,
\end{aligned}
\end{equation}
where $\adj{\op{T}}$, given by
\[
[\adj{\op{T}} \psi](\vect{x}) = \sum_{j=1}^R h_j(\vect{x}) \left\{ \psi(\vect{x}+\vect{\nu}_j) - \psi(\vect{x})\right\},
\]
denotes the adjoint of the operator $\op{T}$ with respect to the bilinear form $(p, \psi) := \sum_{\vect{x}}{p(\vect{x}) \psi(\vect{x})}$.
For a linear reaction network with mass-action kinetics, the corresponding first-order moment equations (the \emph{mean equations}) are closed and linear:
\begin{equation*} 
\ddt \mean{\vect{x}}_t = \sum_{j=1}^R \mean{h_j(\vect{x}_t)} \vect{\nu}_j = \sum_{j=1}^R h_j(\mean{\vect{x}}_t) \vect{\nu}_j = \matr{A}\mean{\vect{x}}_t + \vect{b},
\end{equation*}
where the matrix $\matr{A}$ and the vector $\vect{b}$ are defined by this equation.
If the network is non-linear, the mean equations are not closed. \\


At this point, we turn to considering deterministic dynamics of reaction networks. One possibility to obtain a deterministic description of the dynamics is to start with the continuous-time Markov chain model and take the large volume limit where the reaction rate equations are recovered. To achieve this, we introduce the concentration variable $\vec{\mu} := \vect{X}(t)/V$, where $V$ is the system size. 
Inserting this concentration variable in the corresponding process representation of a continuous-time Markov chain, namely in the random time-change representation of Kurtz\cite{kurtz1980representations} and applying the law of large numbers, one obtains the well-known reaction rate equation (RRE)
\begin{align} \label{eq:RRE}
\ddt  \vec{\mu}(t) =  \sum_{j=1}^{R} \lambda_j(\vec{\mu}(t)) ~\vec{\nu}_j.
\end{align} 
Here, the rate functions $\lambda_j$ are called mass-action propensities for concentrations, and in terms of concentrations for subnet and environmental species, $\vect{\mu} = (\vect{\mu}^{\ts}, \vect{\mu}^{\te})$, they are given by
\[
\lambda_j(\vect{\mu}^{\ts}, \vect{\mu}^{\te}) = c_j \prod_{i=1}^{N} (\mu_i^{\ts})^{s_{ij}} \prod_{l=1}^{M} (\mu_l^{\te})^{s'_{lj}}
\]
with the macroscopic rate constants $c_j$.
Hence, for a reaction system with high copy numbers or high concentrations, the dynamics can be well described by the deterministic dynamics, the RRE.

\subsection{The projection operator formalism} \label{sec:projection_cme}
We use the projection operator formalism for both the stochastic and deterministic description to derive effective equations for the dynamics of the subnet.  
For the stochastic description, we apply the formalism\cite{kuhne1978nakajima} (compare also the original work of Nakajima\cite{nakajima1958quantum}, Zwanzig\cite{zwanzig1960ensemble}, and Mori\cite{mori1965transport}) to derive a generalized master equation. That equation describes the evolution of the marginal probability distribution.

For the CME, a natural definition of a marginalization operator is 
\begin{equation} \label{eq:Marg}
[\op{M}p](\vect{x}^{\ts}) = \sum_{\vect{x}^{\te}}p(\vect{x}^{\ts},\vect{x}^{\te}).
\end{equation}
This operator can be used within the projection operator framework as described in the following.

At first, we observe that the marginalization operator maps the distribution of the joint state space of subnet and environment to the marginal distribution over the subnet state space, whereas the projection operator framework requires a projection operator that maps the space of distributions over the joint space to itself.
Therefore, in addition to the marginalization operator $\op{M}$ defined in \eqref{eq:Marg}, a lifting operator $\op{L}$ acting on distributions $p(\vect{x}^{\ts})$ over the subnet states has to be defined. 
We choose\cite{thomas2012rigorous, venturi2014convolutionless}
\[
[\mathsf{L} p](\vect{x}^{\ts}, \vect{x}^{\te}) = q(\vect{x}^{\te})p(\vect{x}^{\ts})
\]
for some fixed distribution $q$ over the environmental states.

The (time-independent) operator $\op{P} := \op{L} \op{M}$ is a projection, i.e., we have $\op{P}^2 = \op{P}$. Defining the orthogonal projection operator $\op{Q} := \op{I} - \op{P}$, we also have
\[
\op{Q}^2=\op{Q}, \quad \op{P}\op{Q}=\op{Q}\op{P}= 0.
\]

Applying $\op{M}$ (respectively, \op{Q}) to the CME \eqref{eq:cme} and using $\op{P} + \op{Q} = \op{I}$ and $\op{P} = \op{L}\op{M}$, we obtain the two equations
\begin{align}
\frac{\mathrm{d}}{\mathrm{d} t} \op{M}p & = \op{M}\op{T}\op{L}\op{M}p + \op{M}\op{T}\op{Q}p \label{Kol1}, \\  
\frac{\mathrm{d}}{\mathrm{d} t} \op{Q}p & =  \op{Q}\op{T}\op{Q}p + \op{Q}\op{T}\op{L}\op{M}p \label{Kol2}.
\end{align}
Note that the projected distribution $p^{\ts} = \op{M}p$ fully contains the exact subnet dynamics due to the special form of our projection operator.

Formally solving \eqref{Kol2} results in
\begin{align*}
\op{Q} p(t) = e^{t\op{Q}\op{T}} \op{Q}p(0) + \int_{0}^{t} {\mathrm{d} t'} e^{(t-t')\op{Q}\op{T}} \op{Q}\op{T} \op{L} p^{\ts}(t')\, ,
\end{align*}
and inserting the latter into \eqref{Kol1} finally yields
\begin{equation} \label{eq:projCME}
\begin{aligned}
\frac{\mathrm{d}}{\df t} p^{\ts}(t) &= \op{M} \op{T} \op{L} p^{\ts}(t) + \int_{0}^{t} {\df t'} \op{M} \op{T} e^{(t-t')\op{Q}\op{T}} \op{Q}\op{T} \op{L} p^{\ts}(t') \\
&\quad + \op{M} \op{T} e^{t\op{Q}\op{T}} \op{Q}p(0).
\end{aligned} 
\end{equation}
Here, the first term reflects the Markovian part of the dynamics and the second (convolution) term the \emph{memory}. 
The third term represents a \emph{noise} since we assume to have no knowledge about the initial distribution of the environment. The noise term can be made to vanish if we assume that $\op{Q}$ is orthogonal to the initial distribution.

Eq.\,(\ref{eq:projCME}) has a reduced number of dimensions compared to the full network, but the description has not become simpler since the equation still constitutes the exact dynamics of the subnet including the effects of the environment on it -- reflected by the memory and noise terms. To obtain simpler expressions, one needs in general approximations, especially for the memory kernel. However, in this paper we restrict our considerations to networks with mono-molecular within-environment as well as subnet-environment interactions, for which an analytical treatment of prototypical environments is possible. 

We can then decompose the full time-evolution operator $\op{T}$ as $\op{T} = \op{T}_0 + \op{T}_1$, where $\op{T}_0$ includes all linear reactions, and $\op{T}_1$ contains the non-linear subnet-subnet interactions.
Strictly speaking, the operator $\op{T}_1$ acts on distributions over the joint space of subnet and environment. However, in the following we will sometimes implicitly consider it to act on distributions over the subnet state only.

\subsection{Mean equation for projected CME} \label{sec:marg_relations}
In this section, we extract marginal mean equations from the marginal CME \eqref{eq:projCME}.
Starting from \eqref{eq:projCME}, the expectation $\mean{\phi}$ of any function $\phi(\vect{x}^{\ts})$ evolves according to
\begin{equation} \label{eq:cme_marg_moments}
\begin{aligned}
\frac{\df}{\df t} \mean{\phi}_t &= \mean{\op{L}^{\dagger} \op{T}^{\dagger} \op{M}^{\dagger} \phi}_t \\
&\quad + \int_0^t{\df t' \mean{ \op{L}^{\dagger} \op{T}^{\dagger}  \op{Q}^{\dagger} e^{(t-t')\op{T}^{\dagger} \op{Q}^{\dagger}}  \op{T}^{\dagger} \op{M}^{\dagger} \phi }_{t'} } \\
&\quad + \mean{\op{Q}^{\dagger} e^{t \op{T}^{\dagger} \op{Q}^{\dagger}} \op{T}^{\dagger} \op{M}^{\dagger} \phi}_0.
\end{aligned}
\end{equation}
Thus, we need expressions for the adjoints of the operators involved. Let $\psi(\vect{x}^{\ts}, \vect{x}^{\te})$ and $\phi(\vect{x}^{\ts})$ be two arbitrary functions.
A brief computation shows that we have
\[
\begin{aligned}
[\op{M}^{\dagger} \phi](\vect{x}^{\ts}, \vect{x}^{\te}) &= \phi(\vect{x}^{\ts}), \\
[\op{L}^{\dagger}\psi](\vect{x}^{\ts}) &= \sum_{\vect{x}^{\te}} q(\vect{x}^{\te}) \psi(\vect{x}^{\ts}, \vect{x}^{\te})
\end{aligned}
\]
and thus
\[
[\op{P}^{\dagger}\psi](\vect{x}^{\ts}, \vect{x}^{\te}) = \sum_{\vect{\hat{x}}^{\te}} q(\vect{\hat{x}}^{\te}) \psi(\vect{x}^{\ts}, \vect{\hat{x}}^{\te}).
\]
Choosing $\vect{\phi}(\vect{x}^{\ts}) = \vect{x}^{\ts}$, we obtain from \eqref{eq:cme_marg_moments} equations for the mean abundances. To get an explicit expression, we first consider the case of a fully linear network.
For a linear reaction network, all reactions are of one of the two forms
\[
\emptyset \longrightarrow [\cdots], \quad \quad \textrm{X}_i \longrightarrow [\cdots],
\]
where the first form can be seen as a special case of the second form by including an auxiliary species $\textrm{X}_0$
with abundance 1 and replacing the reaction
\[
\emptyset \longrightarrow [\cdots] \quad \textrm{ by } \quad \textrm{X}_0 \longrightarrow \textrm{X}_0 + [\cdots].
\]
We will assume in the following that this rewriting has been performed.

Noting that $\phi$ is linear, we can obtain these equations in a more explicit form by first verifying that each operator involved in \eqref{eq:cme_marg_moments} leaves the space of affine-linear functions invariant. Then \eqref{eq:cme_marg_moments}, which is infinite-dimensional, reduces to a finite-dimensional equation for the coefficients of an affine-linear function.
Thus, let
\[
\psi(\vect{x}^{\ts}, \vect{x}^{\te}) = \vect{w}^{\ts} \vect{x}^{\ts} + \vect{w}^{\te} \vect{x}^{\te} + v,
\quad \phi(\vect{w}^{\ts}) = \vect{w}^{\ts} \vect{x}^{\ts} + v
\]
be general affine-linear functions, where $\vect{w}^{\ts}$ and $\vect{w}^{\te}$ are row vectors and $v$ is a scalar.
As explained in Section~\ref{sec:dynamics}, the first-order moment equations for a linear network are closed. The action of the operator $\adj{\op{T}_0}$ can then be specified via the  matrix
\[
\matr{A} = \begin{bmatrix} \matr{A}^{\tss} & \matr{A}^{\tse} \\ \matr{A}^{\tes} & \matr{A}^{\tee} \end{bmatrix}
\]
as
\[
[\op{T}_0^{\dagger} \psi](\vect{x}^{\ts}, \vect{x}^{\te}) = [\vect{w}^{\ts}, \vect{w}^{\te}] \matr{A} \begin{bmatrix}\vect{x}^{\ts}\\ \vect{x}^{\te}\end{bmatrix}.
\]
Similarly,
\begin{align*}
[\op{M}^{\dagger} \phi](\vect{x}^{\ts}, \vect{x}^{\te}) &= \vect{w}^{\ts} \vect{x}^{\ts} + v,\\
[\op{L}^{\dagger} \psi](\vect{x}^{\ts}) &= \vect{w}^{\ts} \vect{x}^{\ts} + \vect{w}^{\te} \mean{\vect{x}^{\te}}_{q} + v, \\
[\op{P}^{\dagger} \psi](\vect{x}^{\ts}, \vect{x}^{\te}) &= \vect{w}^{\ts} \vect{x}^{\ts} + \vect{w}^{\te} \mean{\vect{x}^{\te}}_{q} + v, \\
[\op{Q}^{\dagger} \psi](\vect{x}^{\ts}, \vect{x}^{\te}) &= \vect{w}^{\te} (\vect{x}^{\te} - \mean{\vect{x}^{\te}}_{q}), \\
[\op{T}_0^{\dagger} \op{Q}^{\dagger} \psi](\vect{x}^{\ts}, \vect{x}^{\te}) &= (\vect{w}^{\te} \matr{A}^{\tes}) \vect{x}^{\ts} + (\vect{w}^{\te} \matr{A}^{\tee}) \vect{x}^{\te}.
\end{align*}
Thus, each operator leaves the space of affine-linear functions invariant. Additionally, the operator $\op{T}_0^{\dagger}$ removes the inhomogeneous term $v$.

We can now evaluate each of the three terms in \eqref{eq:cme_marg_moments} in turn.
For the Markovian part, we obtain
\begin{equation} \label{eq:marg_cme_mean_markovian}
\mean{\op{L}^{\dagger} \op{T}^{\dagger} \op{M}^{\dagger} \vect{\phi}}_t = \mean{\adj{\op{T}_1}\vect{\phi}}_t + \matr{A}^{\tss} \vect{\mu}^{\ts} + \matr{A}^{\tse} \mean{\vect{x}^{\te}}_q.
\end{equation}

To evaluate the memory term, we first note that, because $\adj{\op{M}}\phi$ does not depend on $\vect{x}^{\te}$, we have $\adj{\op{Q}}\adj{\op{T}}\adj{\op{M}}\phi = \adj{\op{Q}}\adj{\op{T}_0}\adj{\op{M}}\phi$. The latter does not depend on $\vect{x}^{\ts}$, so that we obtain $\adj{\op{T}}\adj{\op{Q}}\adj{\op{T}}\adj{\op{M}}\phi = \adj{\op{T}_0}\adj{\op{Q}}\adj{\op{T}_0}\adj{\op{M}}\phi$.
Similarly,  one can verify that that $\adj{\op{T}}\adj{\op{Q}} \phi = \adj{\op{T}_0}\adj{\op{Q}} \phi$.
For the memory term of \eqref{eq:cme_marg_moments}, we note that the coefficients of the linear function $e^{(t-t')\op{T}^{\dagger} \op{Q}^{\dagger}}\op{T}^{\dagger} \op{Q}^{\dagger}\op{T}^{\dagger} \op{M}^{\dagger} \phi = e^{(t-t')\op{T}_0^{\dagger} \op{Q}^{\dagger}}\op{T}_0^{\dagger} \op{Q}^{\dagger}\op{T}_0^{\dagger} \op{M}^{\dagger} \phi$ are the solution of the ODE system
\[
\begin{aligned}
\dot{\vect{w}}^{\ts} &= \vect{w}^{\te} \matr{A}^{\tes}, \\
\dot{\vect{w}}^{\te} &= \vect{w}^{\te} \matr{A}^{\tee}
\end{aligned}
\]
at time $t-t'$ with initial conditions at time 0 given by the coefficients of the linear function $\op{T}_0^{\dagger} \op{Q}^{\dagger}\op{T}_0^{\dagger} \op{M}^{\dagger} \phi$.
Combining all of this, we obtain for the memory term
\begin{align*}
&\int_0^t{\df t' \mean{ \op{L}^{\dagger} \op{T}^{\dagger}  \op{Q}^{\dagger} e^{(t-t')\op{T}^{\dagger} \op{Q}^{\dagger}}  \op{T}^{\dagger} \op{M}^{\dagger} \phi }_{t'} } \\
&=\int_0^t{\df t' \matr{A}^{\tse} e^{(t-t')\matr{A}^{\tee}} \left\{  \matr{A}^{\tes} \vect{\mu}^{\ts}(t') + \matr{A}^{\tee} \mean{\vect{x}^{\te}}_q \right\} }\\
&=\int_0^t{\df t' \matr{A}^{\tse} e^{(t-t')\matr{A}^{\tee}} \matr{A}^{\tes} \vect{\mu}^{\ts}(t')} \\
&\quad + \matr{A}^{\tse}e^{t \matr{A}^{\tee}} \mean{\vect{x}^{\te}}_q - \matr{A}^{\tse} \mean{\vect{x}^{\te}}_q.
\end{align*}

For the noise term in \eqref{eq:cme_marg_moments}, we use the relation  $\op{Q}^{\dagger} e^{t \op{T}^{\dagger} \op{Q}^{\dagger}} = \op{Q}^{\dagger} e^{t \op{Q}^{\dagger} \op{T}^{\dagger} \op{Q}^{\dagger}}$. Since $\op{Q}^{\dagger} \op{T}^{\dagger} \op{Q}^{\dagger} \adj{\op{T}_1} \adj{\op{M}} \phi = 0$,  we have
\[
\op{Q}^{\dagger} e^{t \op{Q}^{\dagger} \op{T}^{\dagger} \op{Q}^{\dagger}} \adj{\op{T}_1} \adj{\op{M}} \phi
= \adj{\op{Q}} \adj{\op{T}_1} \adj{\op{M}} \phi = 0.
\]
We then find
\begin{align*}
\mean{\op{Q}^{\dagger} e^{t \op{Q}^{\dagger} \op{T}^{\dagger} \op{Q}^{\dagger}} \adj{\op{T}} \adj{\op{M}} \phi}_0
&= \mean{\op{Q}^{\dagger} e^{t \op{Q}^{\dagger} \op{T}^{\dagger} \op{Q}^{\dagger}} \adj{\op{T}_0} \adj{\op{M}} \phi}_0 \\
&= \mean{\op{Q}^{\dagger} e^{t \op{Q}^{\dagger} \op{T}_0^{\dagger} \op{Q}^{\dagger}} \adj{\op{T}_0} \adj{\op{M}} \phi}_0 \\
&= 0.
\end{align*}

The full equation for the means thus reads
\begin{equation} \label{eq:marg_mean_eq_linear}
\begin{aligned}
\ddt \mean{\vect{x}^{\ts}}_t &= \mean{\adj{\op{T}_1} \vect{x}^{\ts}}_t + \matr{A}^{\tss}\mean{\vect{x}^{\ts}}_t + \int_0^t{\df t' \matr{A}^{\tse} e^{(t-t')\matr{A}^{\tee}} \matr{A}^{\tes} \mean{\vect{x}^{\ts}}_{t'}} \\
&\quad + \matr{A}^{\tse}e^{t \matr{A}^{\tee}} \mean{\vect{x}^{\te}}_q.
\end{aligned}
\end{equation}
We see that the non-linearity of the subnet enters only in the Markovian part.

At this point, we proceed to relate the projection operator formalism for the CME as given by \eqref{eq:projCME} and the corresponding mean equation (\ref{eq:marg_mean_eq_linear}) to the formalism applied to the ODE model, as given by the reaction rate equation \eqref{eq:RRE}.
In particular, we observe that in the case of linear subnet-environment and intra-environment interactions, the RRE \eqref{eq:RRE} takes the form
\begin{equation} \label{eq:determ_joint}
\begin{aligned}
\ddt \vect{\mu}^{\ts} &= f(\vect{\mu}^{\ts}) + \matr{A}^{\tse} \vect{\mu}^{\te}, \\
\ddt \vect{\mu}^{\te} &= \matr{A}^{\tee}\vect{\mu}^{\te} + \matr{A}^{\tes} \vect{\mu}^{\ts} + \vect{b}^{\te}.
\end{aligned}
\end{equation}
Here, $f$ specifies the subnet-subnet interactions, which can be arbitrary and, in particular, non-linear. Due to the linearity of the remaining reactions, the corresponding dynamics can be represented via (coupling) matrices $\matr{A}$, with entries consisting of rate constants $\{c_j\}$, and the vector $\vect{b}^{\te}$, containing the rate constants of the environmental birth reactions.

Similarly as for the CME, we can derive an effective (projected) subnet description for \eqref{eq:determ_joint}. We solve the equation for $\vect{\mu}^{\te}$ and plug the result into the equation for $\vect{\mu}^{\ts}$. The solution reads
\begin{equation} \label{eq:marg_RRE}
\begin{aligned}
\dot{\vect{\mu}}^{\ts} &= f(\vect{\mu}^{\ts}) + \int_0^t{\df t' \matr{A}^{\tse} e^{(t-t')\matr{A}^{\tee}} (\matr{A}^{\tes} \vect{\mu}^{\ts}(t') + \vect{b}^{\te})} \\
&\quad+ \matr{A}^{\tse}e^{t \matr{A}^{\tee}} \vect{\mu}^{\te}(0).
\end{aligned}
\end{equation}
This is the marginal mean equation as given by \eqref{eq:marg_mean_eq_linear} if we choose the distribution $q$ to have the correct mean, $\mean{\vect{x}^{\te}}_q = \vect{\mu}^{\te}(0)$.
This demonstrates that the marginal CME and the marginal equations obtained from \eqref{eq:determ_joint} are consistent.
Interestingly though, in our derivation above, the Markovian, memory and noise terms in \eqref{eq:cme_marg_moments} and \eqref{eq:marg_mean_eq_linear} do not correspond to each other directly. While the noise term in \eqref{eq:cme_marg_moments} vanishes, the memory term in \eqref{eq:cme_marg_moments} includes $\matr{A}^{\tse}e^{t \matr{A}^{\tee}} \mean{\vect{x}^{\te}}_q$ which is the noise term in \eqref{eq:marg_mean_eq_linear} since it depends on the conditions of the environment which are unknown in general. It also corresponds to the noise term in (\ref{eq:marg_RRE}) if $q$ is chosen as stated above: $\mean{\vect{x}^{\te}}_q = \vect{\mu}^{\te}(0)$.
Also, the noise term in \eqref{eq:cme_marg_moments} vanishes in the first-order moment equations even when the initial distribution $p_0(\vect{x}^{\ts}, \vect{x}^{\te})$ does not factorize into a product $p_0(\vect{x}^{\ts}) p_0(\vect{x}^{\te})$.

Since \eqref{eq:marg_RRE} will be used in Section~\ref{sec:lin_environments} below for the investigation of memory effects in synthetic biology, we want to give some intuition into the different terms of \eqref{eq:marg_RRE}, similarly as for (\ref{eq:projCME}).
The first term $f\left( \vect{\mu}^{\ts} \right)$ describes the dynamics that is generated by all the reactions that have only reactants from inside the subnetwork. We will hence call this term in the following sections the \textit{local dynamics} part.

The second term $\matr{A}^{\tse}e^{t \matr{A}^{\tee}} \, \vect{\mu}^{\te}(0) =: \vect{r}(t)$ is the contribution that originates in the (often unknown) initial condition of the environment.
Since this term cannot be affected by the subnetwork itself and depends solely on the initial conditions, this term is coined \textit{noise} term.

The third term
\begin{align}
\label{eq:memorykernelalone}
\int_0^t \df t' \matr{M}(t-t') \,\vect{\mu}^{\ts}(t')
,\quad \matr{M}(t-t') = \matr{A}^{\tse} e^{(t-t')\matr{A}^{\tee}} \matr{A}^{\tes}
\end{align}
is the \textit{memory term}, where $\matr{M}(t-t')$ denotes the so called \textit{memory kernel} that depends on the dynamics happening inside the environment during the time interval $t-t'$. This term is due to the fact that past states of the subnetwork affected the time evolution of the environment, which now acts back on the subnetwork. The matrices $\matr{A^{\tse}}$ and $\matr{A^{\tes}}$  are the coupling matrices from the environment to the subnetwork and from the subnetwork to the environment, respectively.
In contrast to the noise term, the influence of the memory term on the subnetwork dynamics is therefore predictable upon knowledge of the structure of the environment.

Although in this work we focus on the marginal mean equation as a first measure to quantify the dynamical behavior of the subnetwork, one can similarly derive the marginal second moment equation of the subnetwork for a completely linear reaction network (see Appendix A).



\section{\label{sec:lin_environments}Prototypic Environments with mono-molecular Reactions}

While the full environmental network typically has an unknown and complex structure, the interactions between subnetwork and environmental network can be simple, e.g., the binding of a subnetwork species to an environmental promoter and vice versa, or constant inflow and outflow of species. Proteins can also be bound in more complex topologies like moiety-conserved cycles or temporarily evolve into promiscuous conformers \cite{hofmeyr_metabolic_1986,schwander_synthetic_2016,tokuriki_protein_2009}. Likewise, RNA-folding can lead to transitions between different  transient structures, where only one is functional \cite{semrad_proteins_2011}. 
These parasitic interactions and  the possibly resulting memory terms need to be considered when designing and evaluating synthetic genetic networks. Thus, it is desirable to obtain memory kernels for some prototypical environments. To this purpose, we will analytically derive memory functions for the broad class of chain-like reaction topologies with mono-molecular and linear interactions as depicted in Fig.\,\ref{fig:linchain}. Since in biology almost every reaction is bidirectional (where one direction is conventionally preferred), we include possible reverse reactions in our environments. Additionally, each species has a constant in- and linear outflow due to birth and death reactions. Such a system is sketched in Fig.\,\ref{fig:linchain}, where $\sigma$ is the rate of inflow at each node, $\kappa$ is the decay rate and $\alpha, \beta$ are the rates on the connections along the chain.
Before we derive the corresponding memory functions, we will show that a chain only exposed to constant inflow is (after an initial transient time, which we do not consider) in a steady state and is hence captured by the noise term $r(t)$ that does not affect the memory terms. 
The noise term can be obtained by solving the following system of coupled linear equations:
\begin{align}
x_1 &= \frac{\sigma + x_2 \beta}{\alpha + \kappa}, \nonumber \\
x_m &= \frac{\sigma + \alpha x_{m-1} + \beta x_{m+1}}{\alpha + \beta + \kappa}, \\
x_M &= \frac{\sigma + \alpha x_{M-1}}{\alpha + \beta + \kappa}, \nonumber \\
r(t) &= r = \alpha x_M. \nonumber
\end{align}
Thus, a constant influx to the environmental nodes can  be accounted for  by adding a constant to the noise term.
This implies that the only nontrivial solution ($\alpha \neq 0$) for the noise term is given by $\sigma = \kappa = 0$. Consequently, for this parameter setting the environment is mass conserving.
\subsection{Mapping onto a random walk}
Before calculating the memory function,  we will first show that the considered environment can be mapped onto a random-walk problem. Subsequently, we obtain a return-time distribution of the random walk problem, which is identical to the memory function of the chosen environment.
\begin{figure}
\centering
\includegraphics[width=0.96\linewidth]{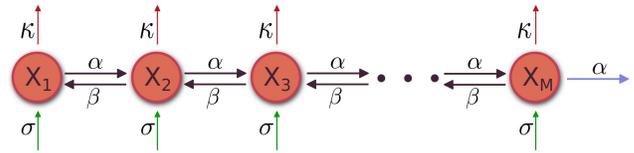}
\caption{Linear chain topology with mono-molecular interactions as considered as generic environment. The rates $\alpha$ and $\beta$ denote the transition rates between the different species $X_i$. $\sigma$ and $\kappa$ are the rates of constant in- and linear outflux. The blue arrow indicates the coupling to the subnetwork. \label{fig:linchain}}
\end{figure}

We explicitly model the interactions between subnetwork and environment. The subnetwork couples to the environment at some arbitrary position $x$ and the environmental chain couples back to the subnetwork at some other position(s) $y_{1/2}$. 
Since in this case the coupling from the subnetwork to the environment consists only of one single reaction, $\matr{A}^{\tes}$ will only have one non-zero element. 
Further, we note that one key feature of a system of mono-molecular reactions is that there is no interaction between different molecules in the network, i.e., the reaction rate of a molecule is independent from that of all the others. 
Hence, inside the environment the mean evolution equation of the system's state is equal to its probability evolution equation. To obtain the general mean equation for the system with $N$ particles, we need to scale up, 
\begin{align}
\mean{\vect{x}}_t = N \vect{p}_t. \nonumber
\end{align}
Since, additionally, for a linear system the mean equation is identical to the deterministic evolution equation -- which we use in our calculations --
it is possible to describe the dynamics of our environmental system by
\begin{align}
\vect{\mu}^{\te} = N \vect{p}^{\te} \nonumber,
\end{align}
where $N$ is the number of particles in the system. 

Additionally, for networks with mono-molecular interactions it was shown that an initial multinomial distribution $M(\vect{x},\vect{p}(0))$ stays multinomial over time\cite{jahnke_solving_2007}. The parameter $\vect{p}$ of this distribution evolves according to the rate equation 
\begin{align}
    p(t) = e^{\matr{B}t}p(0),
    \label{eq:multinom}
\end{align}
where $\matr{B}$ is the deterministic transition rate of the network. In a network with an absorbing state $x_{\textrm{abs}}$ and corresponding $p_{\textrm{abs}}$, the probability to be already absorbed at time $t$ is then given by
\begin{align}
    P_{\textrm{abs}}(t)= 1-M\left(x_{\textrm{abs}}=0,\vect{p}\left(t\right)\right) = p_{\textrm{abs}}(t)\, . \nonumber
\end{align}
Hence, the probability to get absorbed at time $t$ is
\begin{align}
    \frac{\mathrm{d} P_{\textrm{abs}}}{\mathrm{d}t} = \frac{\mathrm{d} p_{\textrm{abs}}}{\mathrm{d}t} = \matr{B}^{\textrm{AN}} \vect{p}(t) 
    \label{eq:absorb}
\end{align}
where $\matr{B}^{\textrm{AN}}$ contains the transition rates from the network into the absorbing state.

Using this features of systems with mono-molecular interactions, we can reinterpret the memory kernel in 
\eqref{eq:memorykernelalone}.
The second part $\matr{A^{\tes}} \vec{\mu}^{\ts}(\tau) \mathrm{d}\tau =: \vec{p}^{\tes}(\tau)$ in \eqref{eq:memorykernelalone} is the concentration that enters the environment during the time interval $\mathrm{d}\tau$ coming from the subnetwork. This initial distribution of one species can be interpreted as a multinomial distribution with $\vect{p} = \vect{e}_i$. In the first part of \eqref{eq:memorykernelalone}, the matrix $\matr{A^{\tse}}$ denotes (similarly to $\matr{B}^{\textrm{AN}}$ in \eqref{eq:absorb}) the rate to be absorbed in the subnetwork. The term $e^{\matr{A^{\tee}} (t - \tau) }$ is the deterministic solution of \eqref{eq:multinom}, where $\matr{A}^\tee$ is the deterministic rate matrix of the environment only. Hence, $\matr{A^{\tse}}e^{\matr{A^{\tee}} (t - \tau) } =: \vec{p}^{\tse}(t - \tau | \tau) $  denotes the rate to leave the environment after a time $t - \tau$ by entering the subnetwork again, given that the environment was entered at time $\tau$.

Naturally, a particle that decays inside the environment will no longer be able to couple back to the subnetwork. Since there is no interaction between the single particles, this directly suggests how linear decay enters the memory terms:
\begin{align}
\vec{p}^{\tse}(t - \tau \mid \tau) = \matr{A^{\tse}} e^{\matr{\bar{A}^{\tee}} (t - \tau) } e^{-\kappa (t-\tau)}
\end{align}
where $\matr{\bar{A}^{\tee}}$ denotes the environmental interactions without decay terms.
We now reinterpret each environmental state (denoted by $X_i$ in Fig.\,\ref{fig:linchain}) as a possible position of a random walker and take each (linear) reaction rate as the corresponding transition probability between the states. As a consequence, to obtain $\matr{A^{\tse}} e^{\matr{A^{\tee}} (t - \tau) } =\vec{p}^{\tse}(t - \tau | \tau) $ we can use a formalism used to analyze random walks with absorbing boundaries to account for the coupling to the subnetwork. Independently from us Stephan et al. \cite{stephan_memory_2018} developed a complementary approach to tackle similar linear environments.


\subsection{Absorption probability for continuous-time random walk \label{sec:randwalk}}
We will now derive a general expression $\vec{p}^{\tse}(t - \tau \mid \tau)$ for any type of a linear random walk with absorbing boundaries.
The problem at hand is a general question of first-return times of a one-dimensional random walk, before coming to more specific situations in later sections. In order to simplify the notation, we will enumerate the states in the chain with $\omega = 1,\dots, \Omega$. As we will show later, it is the most general approach to start with a description that has two absorbing states, namely the states $\omega = 0$ and $\omega = \Omega$. We use the backward master equation to describe the probability $P_{\omega, \chi}(t - \tau)$ of a state starting at $\omega$ to have reached the absorbing state $\chi$ at time $t - \tau$:
\begin{align}
\frac{\mathrm{d}P_{\omega, \chi}(t)}{\mathrm{d}t} = - (\alpha + \beta) P_{\omega, \chi}(t) &+ \alpha P_{\omega+1,\chi}(t) + \beta P_{\omega-1, \chi} (t) \nonumber
\end{align}
where $\alpha$ and $\beta$ are the transition probabilities of one step to the right and left, respectively. 
A problem like this was solved by Heathcote and Moyal \cite{heathcote_random_1959}, where they also assumed a chain with absorbing barriers at $\omega = 0$ and $\omega = \Omega$.
They were able to derive the following equations for the probability to be absorbed at $0$ or $\Omega$ at time $t$:
\begin{widetext}
\begin{align}
\mathcal{P}_{\omega,0}(t)&=\nu ^{-\omega} \sum_{j=0}^{\infty} \int_0^t \tau^{-1} e^{-\mu \tau} \left( \left( 2 j \Omega + \omega \right) I_{2 j \Omega + \omega} - \left( 2 \left( j + 1 \right) \Omega - \omega \right) I_{2\left( j + 1 \right) \Omega - \omega}  \right) d \tau, \nonumber \\
\mathcal{P}_{\omega,\Omega}(t)&=\nu ^{\Omega-\omega} \sum_{j=0}^{\infty} \int_0^t \tau^{-1} e^{-\mu \tau} \left( \left( \left( 2 j + 1 \right) \Omega - \omega \right)
I_{\left( \left( 2 j + 1 \right) \Omega - \omega \right)} - \left( \left( 2 j + 1 \right) \Omega + \omega \right)  I_{\left( \left( 2 j + 1 \right) \Omega + \omega \right)}  \right) d \tau \nonumber
\end{align}
where $\mu = \alpha + \beta$, $\nu = \sqrt[]{\frac{\alpha}{\beta}}$ and $I_x$ is the modified Bessel function of the first kind with argument $ 2 \tau \, \sqrt[]{\alpha \beta} $. Since both final states $0$ and $\Omega$ are absorbing, the probabilities are of cumulative nature and are hence denoted by $\mathcal{P}$ instead of the previously used $P$. We are now interested in the probability to be absorbed exactly after a time-period of $\tau$ which is formally equal to the rate of leaving the environment and thus need the time derivative of these equations:
\begin{equation} \label{eq:twoboundary}
\begin{aligned}
P_{\omega,0}(t)&=\nu ^{-\omega} \sum_{j=0}^{\infty} t^{-1} e^{-\mu t} \left( \left( 2 j \Omega + \omega \right) I_{2 j \Omega + \omega} - \left( 2 \left( j + 1 \right) \Omega - \omega \right) I_{2\left( j + 1 \right) \Omega - \omega}  \right), \\
P_{\omega,\Omega}(t)&=\nu ^{\Omega-\omega} \sum_{j=0}^{\infty}  t^{-1} e^{-\mu t} \left( \left( \left( 2 j + 1 \right) \Omega - \omega \right)
I_{\left( \left( 2 j + 1 \right) \Omega - \omega \right)} - \left( \left( 2 j + 1 \right) \Omega + \omega \right)  I_{\left( \left( 2 j + 1 \right) \Omega + \omega \right)}  \right).
\end{aligned}
\end{equation}
\end{widetext}
Since the environment is finite, we have 
\begin{align}
\int_0^\infty \df t \; (P_{\omega,0}(t) + P_{\omega,\Omega}(t)) = 1, \nonumber
\end{align}
which means that the mass is conserved.

In the following, we will show how we can use the random-walk model to describe different types of generic environments and derive the corresponding memory kernels using the equations that we just derived.


\subsection{Examples of environments and memory kernels}

\subsubsection{One single species}
To give some intuition into the effects of environmental reactions, let us start with one of the simplest examples of a possible environment, namely that of just one species coupled to the subnetwork with rates $\alpha$ and $\beta$ (Fig.\,\ref{fig:context_chain}a). Examples of this type of context effects are e.g. the reversible binding of one subnetwork species to an environmental site. In this case, the only non-zero elements of the coupling matrizes are $A^{\tse}=-A^{\tee} = \beta$ and $A^{\tes} = \alpha$, and the only non-zero element of the memory-kernel \eqref{eq:memorykernelalone} simplifies to
\begin{align}
\label{eq:single_species}
M(\tau) =  \beta e^{-\beta \tau}\,\alpha .
\end{align}
This is the expected exponential probability distribution of any quantity decaying with a linear rate.  

\begin{figure*}
\centering
\includegraphics[width=\linewidth]{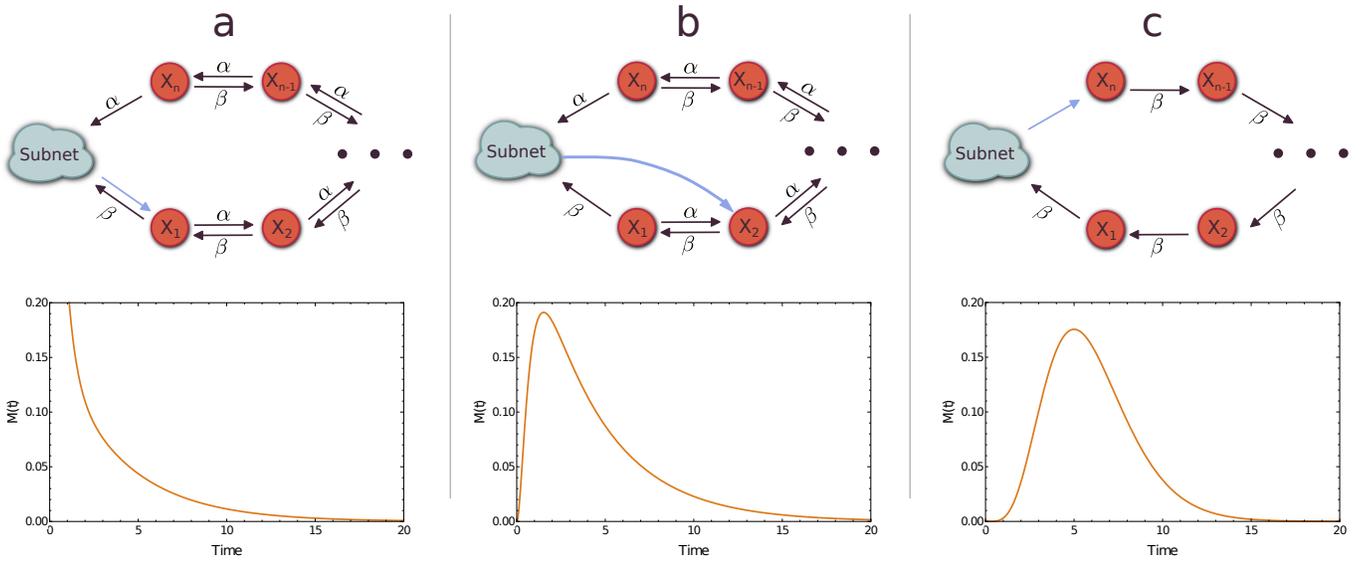}
\caption{ Exemplary sketch of the topology (top) and the corresponding shape of the memory function (bottom) of circular environments as described in Section \ref{sec:cicrular_bi}: (a,b) Bi-directional linear chain with coupling into the environment at (a) the first position (b) some intermediate position (see \eqref{eq:loop}); (c) Directed linear chain (see \eqref{eq:erlang}); Parameters are: $\alpha = \beta = 1$, $\Omega = 6$. Additionally for (b): $\omega = 3$. The three dots denote a arbitrary long continuation of the chain. 
\label{fig:context_loop}}
\end{figure*}

\subsubsection{Chain with two connections to the subnetwork \label{sec:cicrular_bi}}
We now consider the most general configuration where the subnetwork couples to an environmental chain as depicted in Fig.\,\ref{fig:context_loop}a,b. The chain couples back to the subnet and can hence be left by the random walker at both ends.
Examples for this type of environment can, e.g., be found in promiscuous conformation changes of proteins. Here, a protein is only functional in its native state (contained in the subnetwork's description). Due to structural variations like fold transitions, the protein can transform into other dysfunctional states that form a loop-like reaction topology (described by the environment) \cite{tokuriki_protein_2009}.
Thus, we can directly use the sum of Eqs.\,(\ref{eq:twoboundary}) that describes a finite chain with absorbing states at both ends to obtain the memory kernel of such a system:
\begin{align} \label{eq:loop}
M_{\textrm{biChain}}(t) = P_{\omega,0}(t) + P_{\omega,\Omega}(t).
\end{align}
In the given example of the promiscuous conformational changes of proteins, the coupling from the subnetwork to the environment is at the first position of the chain (Fig.\,\ref{fig:context_loop}a), hence $\omega = 1$, but in general there is no constraint on the coupling from the subnetwork to the environment (Fig.\,\ref{fig:context_loop}b).

Enzymatic reaction cycles are often highly directional, which means in our case that we can set $\lim_{\alpha \rightarrow 0}$ (Fig.\,\ref{fig:context_loop}c). Thus, there is no longer any absorption at position $\omega = \Omega$, hence $P_{\omega, \Omega}(t) = 0$ and we obtain:
\begin{align} \label{eq:erlang}
M(t) = \lim_{\alpha \rightarrow 0} P_{\omega, 0}(t) = 
\frac{\beta ^{\omega } e^{-t \beta} t^{ \omega - 1}}{\Gamma (\omega )}.
\end{align}
As expected, this resembles the well known Erlang distribution. 
For a chain length of $\omega = 1$ we obtain the same result as in \eqref{eq:single_species}.

\subsubsection{Linear Chain with one connection to the subnetwork \label{sec:oneconnect}}

\begin{figure*}[!htb]
 \centering
 \includegraphics[width=.96\linewidth]{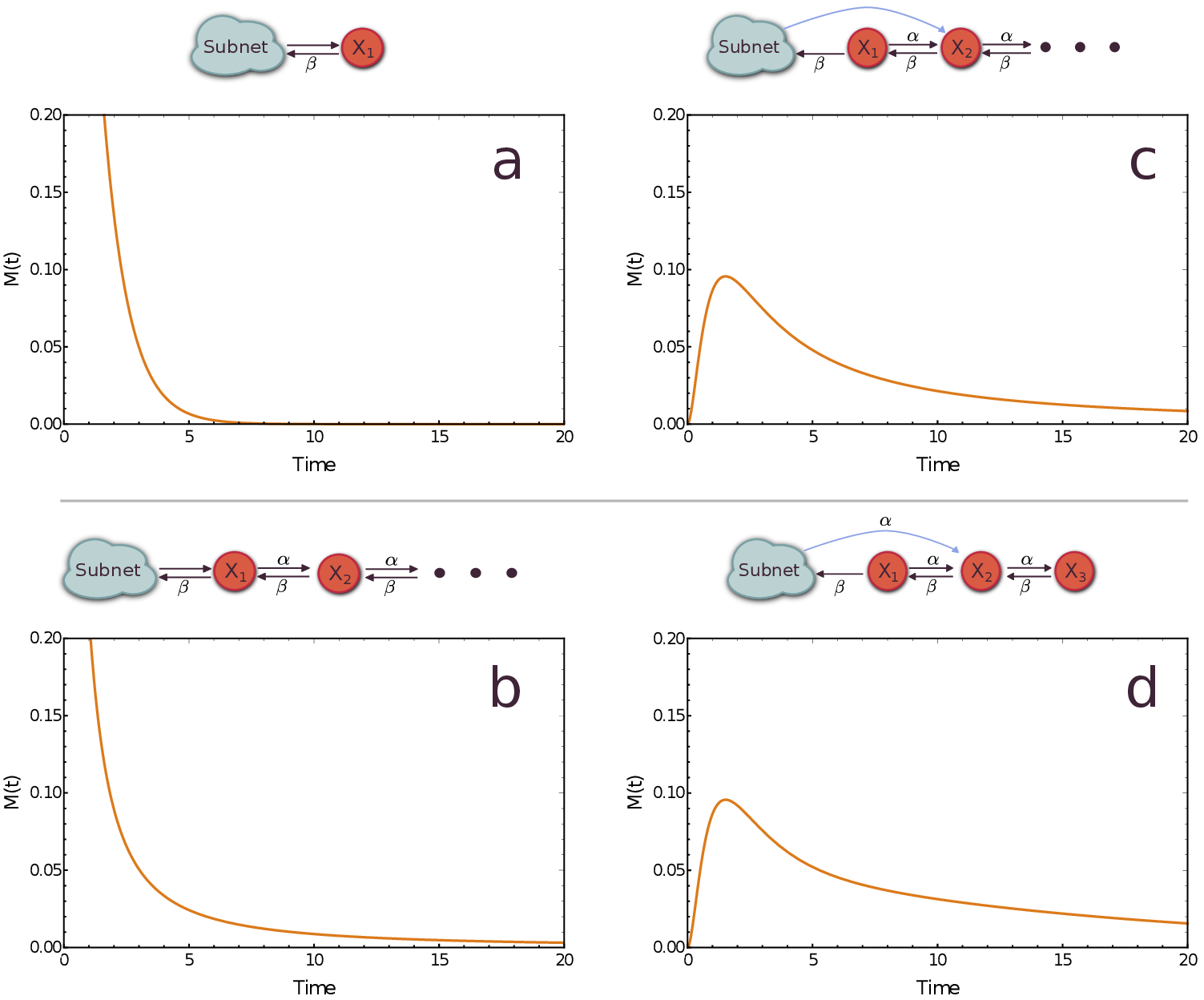}
\caption{
  Sketch of the topology (top) and the corresponding shape of the memory function (bottom) of linear environments with only one coupling to the environment as described in Section \ref{sec:oneconnect}:
(a) Environment with single species (see \eqref{eq:single_species});
(b) Infinite chain of interconnected species, where the coupling between subnet and environment is at the same position (see \eqref{eq:linchain} with $\omega = 1$);
(c) Infinite chain of interconnected species, where the coupling between subnet and environment is at different positions (see \eqref{eq:linchain} with $\omega > 1$);
(d) Finite chain of interconnected species (see \eqref{eq:finitchain});
Parameters are: $\alpha = \beta = 1$. Additionally for (c): $\omega = 3$ and for (d): $\Omega = 6, \omega = 3$.
 \label{fig:context_chain}}
\end{figure*}

Letting $\Omega \rightarrow \infty$ in \eqref{eq:loop},  we minimize the influence of the second absorbing barrier and the probability to be absorbed at $\omega = \Omega$ goes to zero, hence we arrive at the description of some environment with only one coupling to the subnetwork. 
In this case, the probability to be absorbed at $\omega = 0$ at time $t$ simplifies to \cite{heathcote_random_1959}
\begin{widetext}
\begin{align} \label{eq:linchain}
\lim_{\Omega \rightarrow \infty}P_{\omega,0}(t)&= \lim_{\Omega \rightarrow \infty} \nu ^{-\omega} \sum_{j=0}^{\infty} t^{-1} e^{-\mu t} \left( \left( 2 j \Omega + \omega \right) I_{2 j \Omega + \omega} - \left( 2 \left( j + 1 \right) \Omega - \omega \right) I_{2\left( j + 1 \right) \Omega - \omega}  \right) \nonumber \\
&= \omega  \nu ^{-\omega} t^{-1} e^{-\mu t} I_{\omega}.
\end{align}
\end{widetext}
In comparison to the first example of a single-species environment (Fig.\,\ref{fig:context_chain}a), in a longer chain more than one species takes part in buffering species from the subnetwork (Fig.\,\ref{fig:context_chain}b). Hence, the memory function decays slower.

Similar as for the double ended chain, a coupling into the environment at some intermediate position will lead to non-monotonous memory functions (Fig.\,\ref{fig:context_chain}c).

While infinite reaction chains might be a good approximation for  long reaction cascades, chains with a small number of species are in general more relevant. Therefore, in the following, we consider reaction chains of finite length 
with one absorbing state at $\omega = 0$ and one reflecting state at $\omega = \Omega$ (e.g., see Fig.\,\ref{fig:context_chain}d), as e.g. depicted in  Fig.\,\ref{fig:context_chain}d with $\alpha = \beta$.

\begin{figure*}
\centering
\includegraphics[width=.8\linewidth]{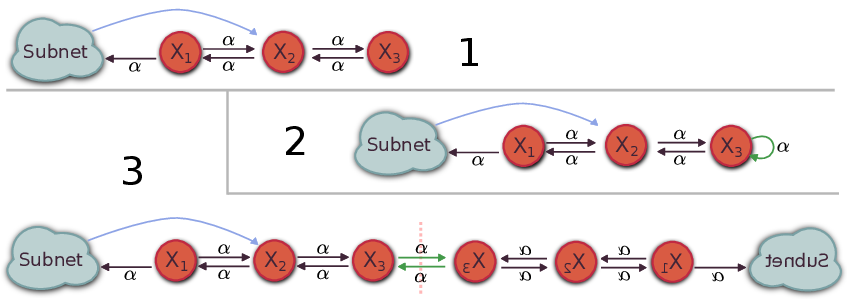}
\caption{ Sketch for the equivalent transformation of a finite chain to a two-ended chain. (1) The initial finite chain we want do describe is extended by a self-referring loop reaction (2, green). This added reaction can be considered as a transition reaction between two equivalent finite chains, that now form a two-ended network topology (3). The ``mirror'' axes is depicted as a pink dashed line.
\label{fig:chain_mirror}}
\end{figure*}

At each position in the chain a particle can move to the left and to the right with a summed propensity of $\alpha + \beta$. The last species (e.g. $X_3$ in the figure) can only move to the left with summed propensity $\beta$. Nevertheless, without changing the dynamics in the system we can add a self-referring loop with rate $\alpha$ to $X_3$. This self-referring loop can also be considered as a connection between two different $X_3$ on two different, but interconnected finite chains (see Fig.\,\ref{fig:chain_mirror}).
Hence, the dynamics of a particle on a chain with reflecting boundary condition is equivalent to the movement of a particle on a chain of length $2 \Omega + 1$ where the absorption rate of the particle starting on position $\omega$ is given by

\begin{align} \label{eq:finitchain}
P_{\textrm{finiteChain} \Omega}(t) = \hat{P}_{\omega, 0}(t) + \hat{P}_{\omega, 2 \Omega + 1}(t),
\end{align}
where $\hat{P}$ denotes (just like in Eq. \eqref{eq:twoboundary}) the rate of leaving the environment, but now with chain length $2 \Omega +1$.

\section{\label{sec:examples}Applications}
\subsection{Example 1: The effect of retroactivity on the repressilator}
\emph{Retroactivity}\\

Del Vecchio et al. describe a general engineering issue that arises as soon as different subsystems (modules) are interconnected to a large device \cite{del_vecchio_modular_2008}. They define \textit{retroactivity} as a signal that travels back from a downstream module to an upstream module and derive a formula to quantify the effect of this retroactivity. A basic assumption of their derivation is the timescale separation between a fast downstream and a slow upstream module. The projection operator framework provides a more general method to analyze the impact of retroactivity on the performance of a genetic module, without relying on timescale separation. Having obtained memory kernels for different prototypic environments, we now give some examples of how context effects can drastically change the behavior of genetic devices. 

Del Vecchio et al. demonstrate the effect of retroactivity on an oscillating input protein signal $k(t)$ that is coupled to a downstream promoter binding region. In insulation, the system shows the desired oscillation, whereas, as soon as interconnected to the fast promoter, the oscillation is damped. In their study, $k(t)$ was used as a generic output signal of an upstream oscillating device. In the general case, the complete functioning of the oscillator will be deteriorated by the coupling to a downstream module. Using the example of the famous repressilator, we will explicitly model the system of interest in terms of reactions that result in an oscillating signal, and we will investigate the effect of retroactivity.

\medskip

\emph{Repressilator}\\

\begin{figure*}
  \centering
  \includegraphics[width=.9\linewidth]{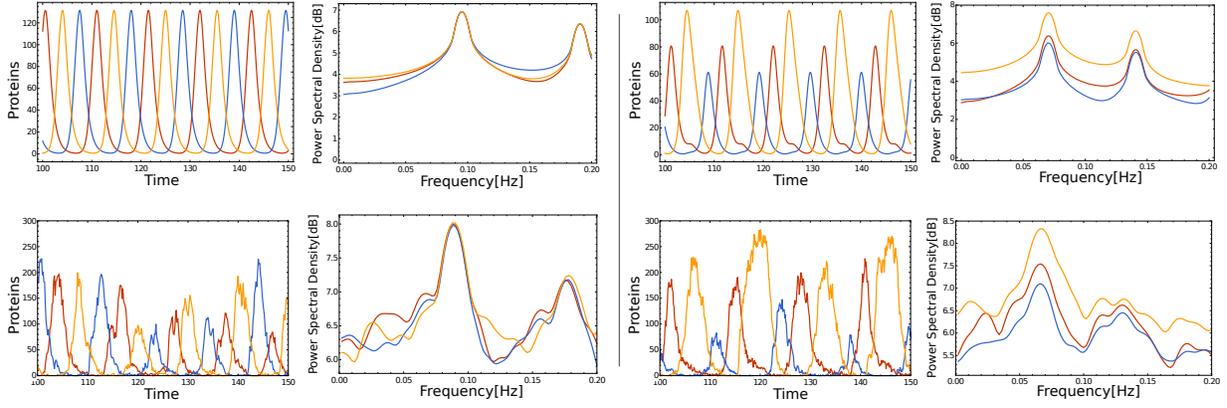}
  \caption{Comparison of deterministic solution (top) and stochastic simulation (bottom) of the described repressilator system, where red, blue and yellow denote $P_A$, $P_B$ and $P_C$, respectively. (left) The unmodified repressilator as given by \eqref{eq:rep}. (right) The repressilator with one protein coupled to an environment as described in \eqref{eq:rep_delay}. Parameters taken from Elowitz and Leibler\cite{elowitz2000synthetic}: $K_p = T = 5$, $K_m = 1$, $n = 2.1$, $K_b = 1$, $\gamma = 250$. Additionally for (right) taken from \cite{del_vecchio_modular_2008}: $k_{\textrm{on}} = 1000$, $k_{\textrm{off}} = 100$.}
  \label{fig:rep_prom}
\end{figure*}

The repressilator is a synthetic genetic regulatory network of three genes and the corresponding three proteins, where each protein represses another gene. 
The resulting interaction can be described by the following equations:
\begin{equation} \label{eq:rep}
\begin{aligned} 
\frac{d m_i}{dt} &= -K_m m_i + \frac{\gamma}{1 + K_b p_j^n}, \\
\frac{dp_i}{dt} &= -K_p p_i + T m_i,
\end{aligned}
\end{equation}
where $m_i$ denotes the concentration of the three different mRNA and the $p_i$ of the proteins. The degradation rates of mRNA and proteins are given by $K_m$ and $K_p$, respectively. $T$ is the translation rate, $\gamma$, $n$ and $K_b$ are the Hill parameters. This corresponds to the following reaction scheme:
\begin{equation} \label{eq:reactionsetOne}
\begin{aligned}
\emptyset &\xrightarrow[]{\mathmakebox[2cm]{\frac{\gamma}{1 + K_b p_j^n}}} M_i, \\
M_i &\xrightarrow[]{\mathmakebox[2cm]{K_m}} \emptyset, \\
M_i &\xrightarrow[]{\mathmakebox[2cm]{T}} M_i + P_i, \\
P_i &\xrightarrow[]{\mathmakebox[2cm]{K_p}} \emptyset. 
\end{aligned}
\end{equation}
Using the parameter set given by Elowitz and Leibler \cite{elowitz2000synthetic}, in the deterministic, as well as in the stochastic simulation one observes a uniform oscillation in the concentrations of the different proteins (see Fig.\,\ref{fig:rep_prom}\,(left)). \\

\medskip

\emph{Retroactivity due to promoter binding}\\

Using the repressilator, we can now build the system used by Del Veccio et al. where we, however, explicitly model the oscillating upstream module.
One of the oscillating proteins of the repressilator is supposed to drive a downstream module by reversibly binding to its promoter. To account for this binding, we add to the reaction equations \eqref{eq:reactionsetOne} a reversible binding reaction for protein $A$:
\begin{align}
P_A  &\xrightleftharpoons[k_{\textrm{off}}]{\mathmakebox[2cm]{k_{\textrm{on}}}} C, \nonumber
\end{align}
where $C$ is the complex protein promoter. Using the results from \eqref{eq:single_species}, this leads to a modified set of differential equations:
\begin{align} \label{eq:rep_delay}
\dot{m_i} &= -K_m m_i + \frac{\gamma}{1 + K_b p_j^n}, \nonumber \\
\dot{p_i} &= -K_p p_i + T m_i \\ 
&\quad + \delta_{Ai} \left( \int_0^t k_{\textrm{off}} e^{-k_{\textrm{off}} (t - \tau) }  k_{\textrm{on}} p_i(\tau) d\tau - p_i k_{\textrm{on}}  \right) \nonumber.
\end{align}
For the isolated repressilator, one expects a driving frequency of the promoter analogous to the one depicted in Fig.\,\ref{fig:rep_prom}\,(left). Nevertheless -- as shown by solving \eqref{eq:rep_delay} as well as by stochastic simulations (Fig.\,\ref{fig:rep_prom}\,(right)) -- due to the reversible binding of one of the proteins the oscillation frequency of the protein concentration is vastly reduced. Additionally, the amplitudes of the protein concentrations are changed due to the coupling to the promoter. 

This simple example highlights the need to account for context effects when designing synthetic genetic modules, especially when time-scale separation is not possible.

\subsection{Example 2: Context effects due to promiscuity}
We now want to proceed to a more sophisticated example of a rather counterintuitive result of context effects. We hereby build on a study by Bratsun et al. \cite{bratsun_delay-induced_2005}. Based on the observation that in gene regulatory networks there is often a vast separation of timescales between different types of reactions, the authors present different gene regulation modules that show oscillations induced by delayed reactions. In the following, we modify their example of a ``negative feedback with dimerization'' by adding a constant inflow of $X$. Additionally, we motivate how the context of downstream modules can cause a delay.

\begin{figure*}
  \centering
  \includegraphics[width=.75\linewidth]{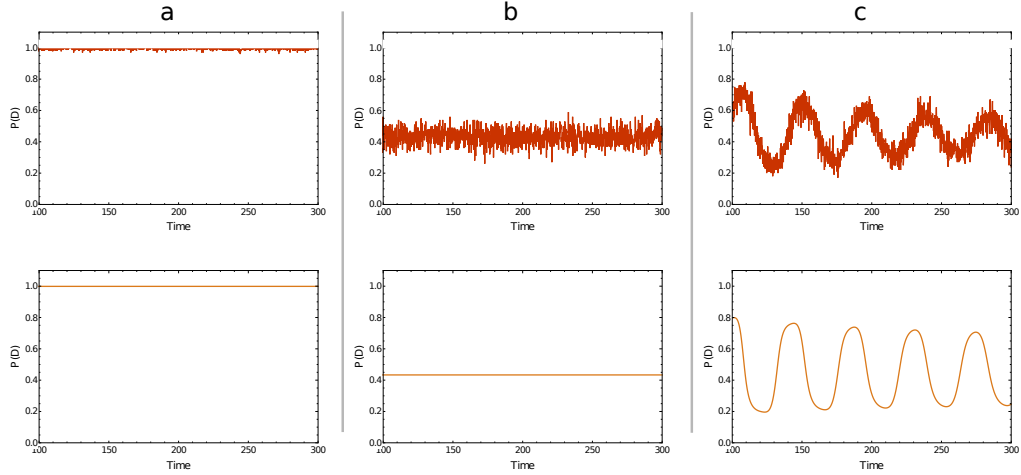}
  \caption{Comparison of deterministic solution (bottom) and stochastic simulation (averaged over 100 runs) (top) for the probability of an active promoter for the gene-regulation module described in Example 2. (a) The unmodified  module without feedback; (b) the module with immediate feedback; (c) the module with delayed feedback. Parameters: $\delta_{\textrm{in}}= 2, \delta_{\textrm{out}}= 4, k_{\textrm{on}}=1000, k_{\textrm{off}}= 100, k_{\textrm{mon}}=1000, k_{\textrm{dim}}= 200, \gamma= 70, \tau = 20$.}
\label{fig:oszi}
\end{figure*}

A common function of synthetic genetic modules is the activation and repression of downstream genes. This is often achieved in two steps, where in the first step a specific monomer has to dimerize before it can bind to an downstream promoter. Denoting the monomer with $X$, the dimer with $X_2$ and the activated and inactivated promoter with $D$ and $D^*$, respectively, the following reactions describe a repression of the promoter  $D$ by $X_2$:
\begin{equation} \label{eq:dimer}
\begin{aligned} 
X + X &\xleftrightharpoons[k_{\textrm{dim}}]{\mathmakebox[2cm]{k_{\textrm{mon}}}} X_2,  \\
X_2 + D &\xleftrightharpoons[k_{\textrm{off}}]{\mathmakebox[2cm]{k_{\textrm{on}}}} D^*, \\
X &\xleftrightharpoons[\delta_{\textrm{out}}]{\mathmakebox[2cm]{\delta_{\textrm{in}}}} \emptyset. 
\end{aligned}
\end{equation}
Here, the relevant output parameter is the probability of an activated/inactivated promoter. The corresponding ODEs read
\begin{equation} \label{eq:ODEDimer}
\begin{aligned}
\dot{x}(t) =&\,\delta_{\textrm{in}}-\delta_{\textrm{out}}\,x(t)-2 k_{\textrm{dim}} x(t)^2+ 2 k_{\textrm{mon}} x_2(t), \\
\dot{x}_2(t) =&- k_{\textrm{on}} d(t) x_2(t)+ k_{\textrm{off}} d^*(t) + k_{\textrm{dim}} x(t)^2- k_{\textrm{mon}} x_2(t), \\ 
\dot{d}(t) =& - k_{\textrm{off}} d(t) x_2(t) + k_{\textrm{on}} d^*(t), \\ 
\dot{d}^*(t)=&\,k_{\textrm{off}} d(t) x_2(t) - k_{\textrm{on}} d^*(t).
\end{aligned}
\end{equation}
Analyzed in isolation, this results in a constant probability for an active promoter that depends on the chosen parameters:
\begin{equation}
P(d) =  \frac{\delta_{\textrm{out}}^2\,k_{\textrm{on}}\,k_{\textrm{mon}}}{\delta_{\textrm{out}}^2\,k_{\textrm{on}}\,k_{\textrm{mon}} + k_{\textrm{dim}}\,k_{\textrm{off}}}.
\end{equation}
An example simulation as well as a numerical solution of the ODE system is shown in Fig.\,\ref{fig:oszi}a.

In a synthetic genetic network, the activated promoter will be used to trigger the operation of downstream modules. Let us assume that there is a downstream source of protein $\hat{X}$ that has -- due to e.g. promiscuity -- properties similar to the $X$ protein. To account for this, and assuming that expression of $\hat{X}$ occurs soon after activation of the promoter $D$, one would add the following reaction to \eqref{eq:dimer}:
\begin{align}
D &\xrightarrow[]{\mathmakebox[2cm]{\gamma}} D + X
\end{align}
and likewise extend the first ODE in \eqref{eq:ODEDimer} that describes the time evolution of $X$:
\begin{align*}
\dot{x}(t)= \,& \delta_{\textrm{in}} -\delta_{\textrm{out}}\,x(t)-2 k_{\textrm{dim}} x(t)^2 +
\\ &2 k_{\textrm{mon}} x_2(t) + \gamma d(t).
\end{align*}
The resulting behavior depicted in Fig.\,\ref{fig:oszi}b essentially means that the feedback due to the interaction between downstream and upstream modules leads to a shift in the probability for an activated promoter. However, when we take into account the time delay between activation of $D$ and expression of $\hat X$, this result becomes modified. Such a time delay becomes significant if the number of modules between the dimerization module and the one expressing $\hat{X}$ is large, or if there are some slow processes between the activation of $D$ and the expression of $\hat{X}$. We account for this by writing:
\begin{align*}
D &\xrightarrow[]{\mathmakebox[2cm]{\gamma}} D + Z_1,\\
Z_1 &\xrightarrow[]{\mathmakebox[2cm]{\gamma}} Z_2,\\
Z_2 &\xrightarrow[]{\mathmakebox[2cm]{\gamma}} Z_3,\\
&\vdots  \\
Z_n &\xrightarrow[]{\mathmakebox[2cm]{\gamma}} X,
\end{align*}
where $Z_i$ are the intermediate steps before $\hat{X}$ is expressed. The chain of $Z_i$ reactions forms an environment similar to the directed chain described under subsection \ref{sec:oneconnect}.
There, we have shown that $M(t-\tau)$ is an Erlang distribution as given by \eqref{eq:erlang}. We thus need to include this memory effect in the equation for $\dot{x}$. We will do this by using a simplification: if we keep the mean $\frac{\omega}{\alpha}$ fixed and increase the size of the chain, the variance decreases with $\frac{1}{\omega}$, and we can simplify:
\begin{align} \label{eq:delta}
\left. \lim_{\omega \rightarrow \infty} M(t-\tau) \right|_{\frac{\omega}{\alpha} = \textrm{const}} = \nonumber \\
\left. \lim_{\omega \rightarrow \infty} \frac{\beta ^{\omega } e^{-(t-\tau) \beta} (t-\tau)^{ \omega - 1}}{\Gamma (\omega )} \right|_{\frac{\omega}{\alpha} = \textrm{const}} = \nonumber \\
\delta \left( t - \tau - \frac{\omega}{\alpha} \right). 
\end{align}
As an approximation, we can hence treat the environment as a memory function that has the form of a time-delayed delta distribution.
Using this result we are now in the position to give an approximate result for the effects of the environment on the probability for an activated promoter. Taking into account this simple form of the memory effect, the equation for $\dot{x}$ now takes the form
\begin{align}
\dot{x}(t)=& \,\delta_{\textrm{in}} -\delta_{\textrm{out}}\,x(t)-2 k_{\textrm{dim}} x(t)^2+ \nonumber \\ 
&2 k_{\textrm{mon}} x_2(t)+ \int_0^t M(t-\tau) d(\tau) d\tau \nonumber \\
=& \,\delta_{\textrm{in}}-\delta_{\textrm{out}}\,x(t)-2 k_{\textrm{dim}} x(t)^2+ \nonumber \\ 
&2 k_{\textrm{mon}} x_2(t) + \gamma \delta(t - \tau),   \nonumber 
\end{align}
where we set $M(t-\tau) = \delta(t-\tau)$ as given by \eqref{eq:delta}.  
In Fig.\,\ref{fig:oszi}c we compare the stochastic simulation and the approximate deterministic solution of the self-repression circuit. Due to the context effects the probability for the promoter to be activated now oscillates in time. As we have shown, this effect can only be observed if the memory effect of the environment is considered.


\section{\label{sec:conclusion}Conclusion}

In this article, we tackled the problem of context effects of synthetic genetic networks that are embedded in a cellular environment. To this purpose, we exploited the projection operator formalism to obtain  an effective description of the target network that includes the environmental network via memory terms. In contrast to Rubin et al. who applied the projection framework to the chemical Langevin equation, we applied it directly to the chemical master equation to obtain the exact Nakajima-Zwanzig equation. For linear networks we showed that the marginal RRE coincides with the first moment equation corresponding to the marginal/projected CME. This was, to the best of our knowledge, not done before. 

Subsequently, we analyzed the memory terms of possible linear environments. 
While previous studies often assumed specific knowledge about the environment, to meet the requirements of synthetic biology, our goal was to avoid assuming specific knowledge of the environmental structure. Therefore, we considered several prototypic environments and deduced the memory kernels induced by them. In particular, we focused on environmental conversion chains of different lengths and with connections to the subnetwork. Such reaction chains are found in a lot of scenarios of cellular environments. By exploiting the mapping between conversion chains and one-dimensional random walks, we found analytical expressions for the corresponding memory kernels. 

Using the example of the repressilator, we applied our results and illustrated how the contributions of the context can deteriorate the system's performance. By coupling one protein species to an environmental promoter, we observed a noteworthy change in the oscillation frequency. Additionally, we used a simple self-repression module to demonstrate how memory terms can lead to counterintuitive phenomena like oscillations that can not be detected with conventional analysis strategies. 

Chain-like environmental cellular reaction structures occur in many cases and most delay distributions can be approximated with unimolecular stochastic reaction networks \cite{briat_ergodicity_2018}. Nevertheless, a reduction to only linear reactions is always an oversimplification, even though it may be a good approximation close to a steady state. Our results pose as a starting point for further analysis of non-linear environmental structures and more generic environments in the context of synthetic biology.


\begin{acknowledgments}
This work was supported by the Landesoffensive für wissenschaftliche Exzellenz (LOEWE; initiative to increase research excellence in the state of Hessen, Germany) as part of the LOEWE Schwerpunkt CompuGene.
\end{acknowledgments}

\appendix
\setcounter{equation}{0}
\renewcommand{\theequation}{A.\arabic{equation}}
\section*{Appendix A}
Subsequently, we provide the derivation of the marginal second-order moment equation for the subnetwork species of a \emph{completely linear} reaction network. Similarly to the derivation of the marginal mean equation in the main text we start to characterize the action of $\adj{\op{T}}$ on quadratic functions. In particular, we define
\begin{align*}
    \psi(\vec{x}) := \vect{x} \otimes \vect{x},
\end{align*}
where $\otimes$ denotes the Kronecker product and we are interested in 
\begin{align*}
    \frac{\mathrm{d}}{\mathrm{d}t} \mean{\psi}_t = \mean{\adj{\op{T}} \psi}_t,
\end{align*}
which is the equation for the second moments of the full network
\begin{align} \label{full_second_moments}
    \frac{\mathrm{d}}{\mathrm{d}t} & \mean{\psi(\vect{X})}_t \equiv \frac{\mathrm{d}}{\mathrm{d}t}  \mean{\vect{X} \otimes \vect{X}}_t= \sum_j \mean{\vect{X} h_j(\vect{X})}_t \otimes \vect{\nu}_j \nonumber \\
    & + \sum_j \vect{\nu}_j \otimes \mean{\vect{X} h_j(\vect{X})}_t + \sum_j (\vect{\nu}_j \otimes \vect{\nu}_j) \, h_j(\mean{\vect{X}}_t),
\end{align}
where we used the linearity of the network in the last summand.
As the resulting moment equations are closed and linear, we do not follow the operator formalism but rather apply the more direct approach of partly solving the joint system explicitly to obtain a reduced set of equations for the moments of the subnetwork. For that purpose we express (\ref{full_second_moments}) and the first order dynamics as  
\begin{align} \label{linear_equ_secMom}
    \frac{\mathrm{d}}{\mathrm{d}t} 
    \begin{bmatrix}
      \mean{\vect{X}}_t \\
        \mean{\vect{X} \otimes \vect{X}}_t
    \end{bmatrix} = 
    \begin{pmatrix}
        \matr{A} & \matr{0} \\
        \matr{B} & \matr{C}
    \end{pmatrix}
    \begin{bmatrix}
        \mean{\vect{X}}_t \\
                \mean{\vect{X} \otimes \vect{X}}_t
    \end{bmatrix},
\end{align}
where the first line corresponds to the mean equations determined by the block matrix $\matr{A}$ given by the RRE. We proceed to characterize the matrices $\matr{B}$ and $\matr{C}$. 

Therefore we exploit linearity to write $h_j(\vect{x}) = (\nabla h_j) \vect{x}$
where all entries of the gradients of the propensities are zero except one that is equal to the rate constant of the corresponding reaction. Thus, with 
\begin{align*}
    \matr{B} := \Bigg[ \sum_j 
    \left(\begin{bmatrix} 
    \vect{\nu}_j^\ts \\ 
    \vect{\nu}_j^\te 
    \end{bmatrix} 
    \otimes 
    \begin{bmatrix} 
    \vect{\nu}_j^\ts \\
    \vect{\nu}_j^\te
    \end{bmatrix} \right) 
    \begin{pmatrix}
    \nabla h_j^\ts, & \nabla h_j^\te
    \end{pmatrix}
    \Bigg],
\end{align*}
we can then write the last term of (\ref{full_second_moments}) as
\begin{align*}
\sum_j (\vect{\nu}_j \otimes \vect{\nu}_j) \, h_j(\mean{\vect{X}}_t) = 
    \matr{B} 
    \begin{bmatrix}
    \mean{\vect{X}^\ts}_t \\
    \mean{\vect{X}^\te}_t
    \end{bmatrix} .
\end{align*}
Whereas, for the first and second summand of (\ref{full_second_moments}) captured by $\matr{C}$, we use the following identities. For column vectors $\vect{x}, \vect{v}, \vect{h}$ of proper dimension, we have
\begin{align*} 
    \vect{x} \adj{\vect{x}} \vect{h} \otimes \vect{v} &= (\matr{I} \otimes \vect{v} \otimes \adj{\vect{h}})\, \text{vec}(\vect{x} \adj{\vect{x}}),\\
    \vect{v} \otimes \vect{x} \adj{\vect{x}} \vect{h} &= (\vect{v} \otimes \matr{I} \otimes \adj{\vect{h}})\, \text{vec}(\vect{x} \adj{\vect{x}}).
\end{align*}
With this, the remaining matrix $\matr{C}$ reads
\begin{align*}
    \matr{C} = 
    \sum_j & \left\{
    \begin{pmatrix}
    \matr{I}_N & \matr{0} \\
    \matr{0} & \matr{I}_M
    \end{pmatrix}
    \otimes 
    \begin{pmatrix}
    \vect{\nu}_j^\ts \\
    \vect{\nu}_j^\te
    \end{pmatrix} 
    +
    \begin{pmatrix}
    \vect{\nu}_j^\ts \\
    \vect{\nu}_j^\te
    \end{pmatrix} 
    \otimes
    \begin{pmatrix}
    \matr{I}_N & \matr{0} \\
    \matr{0} & \matr{I}_M
    \end{pmatrix}
    \right\} \\
    & \otimes 
    \begin{pmatrix}
    \nabla h_j^\ts, & \nabla h_j^\te
    \end{pmatrix}.
\end{align*}
Inserting  $\matr{A},\matr{B}$ and $\matr{C}$ in  \eqref{linear_equ_secMom} and interpreting those components of the differential equation involving environmental variables as non-autonomous equations, we can then solve them explicitly as a function of the variables involving only subnetwork variables. By inserting those expressions into the remaining differential equations involving only subnetwork variables, one obtains an autonomous integro-differential equation of the Mori-Zwanzig type describing the first and second order dynamics of the subnetwork. As the calculation is straightforward and the resulting equations are rather lengthy, we omit the explicit expressions.

\bibliography{references}
    
\end{document}